\documentclass[12pt]{article}
\usepackage{amssymb,amsmath,epsfig}
\begin{document}
\title{\bf Exact Solutions of Bianchi Types $I$ and $V$ Models in $f(R,T)$ Gravity}

\author{M. Farasat
Shamir\thanks{farasat.shamir@nu.edu.pk}, Adil Jhangeer
\thanks{adil.jahangeer@nu.edu.pk} and Akhlaq Ahmad Bhatti \thanks{akhlaq.ahmad@nu.edu.pk} \\\\  Department of
Sciences \& Humanities, \\National University of Computer \&
Emerging Sciences,\\ Lahore Campus, Pakistan.\\ Tel:
92-42-111-128-128(Ext.229), Fax: 92-42-35165232.}

\date{}

\maketitle
\begin{abstract}
This paper is devoted to investigate the exact solutions of
Bianchi types $I$ and $V$ spacetimes in the context of $f(R,T)$
gravity \cite{fRT1}. For this purpose, we find two exact solutions
in each case by using assumption of constant deceleration
parameter and the variation law of Hubble parameter. The obtained
solutions correspond to two different models of this universe. The
physical behavior of these models is also discussed.
\end{abstract}


\section{Introduction}

The most popular issue in the modern day cosmology is the current
expansion of universe. It is now evident from observational and
theoretical facts that our universe is in the phase of accelerated
expansion \cite{acc1}-\cite{acc9}. The phenomenon of dark energy and
dark matter is another topic of discussion \cite{de1}-\cite{de8}. It
was Einstein who first gave the concept of dark energy and
introduced the small positive cosmological constant. But after that,
he remarked it as the biggest mistake in his life. However, it is
now thought that the cosmological constant may be a suitable
candidate for dark energy. Another proposal to justify the current
expansion of universe comes from modified or alternative theories of
gravity. $f(T)$ theory of gravity is one such example which has been
recently developed. This theory is a generalized version of
teleparallel gravity in which Weitzenb\"{o}ck connection is used
instead of Levi-Civita connection. The interesting feature of this
theory is that it may explain the current acceleration without
involving dark energy. A considerable amount of work has been done
in this theory so far \cite{ft}. Another interesting modified theory
is $f(R)$ theory of gravity. In this theory, a general function of
Ricci scalar is used in standard Einstein-Hilbert lagrangian. Some
review articles \cite{rev} can be helpful to understand the theory.

Many authors have investigated $f(R)$ gravity in different
contexts. Spherically symmetric symmetric solutions are most
commonly studied solutions due to their closeness to the nature.
Multam$\ddot{a}$ki and Vilja \cite{fr1} explored vacuum and
perfect fluid solutions of spherically symmetric spacetime in
metric version of this theory. They used the assumption of
constant scalar curvature and found that the solutions
corresponded to the already existing solutions in general
relativity (GR). Noether symmetries have been used by Capozziello
et al. \cite{fr2} to study spherically symmetric solutions in
$f(R)$ gravity. Similarly many interesting results have been found
using spherical symmetry in $f(R)$ gravity \cite{fr3}.
Cylindrically symmetric vacuum and non-vacuum solutions has also
been explored in this theory \cite{cylndr}. Sharif and Shamir
\cite{me1}. found plane symmetric solutions. The same authors
\cite{me2} discussed the solutions of Bianchi types $I$ and $V$
cosmologies for vacuum and non-vacuum cases. We \cite{me3}
calculated conserved quantities in $f(R)$ gravity via Noether
symmetry approach.

In a recent paper \cite{fRT1}, Harko et al. proposed a new
generalized theory known as $f(R,T)$ gravity. In this theory,
gravitational Lagrangian involves an arbitrary function of the
scalar curvature $R$ and the trace of the energy-momentum tensor $T
$. Myrzakulov \cite{fRT2} discussed $f(R,T)$ gravity in which he
explicitly presented point like Lagrangians. The exact solutions of
$f(R,T)$ field equations for locally rotationally symmetric Bianchi
type I spacetime has been reported by Adhav \cite{fRT3}.  Sharif and
Zubair \cite{fRT5} discussed the laws of thermodynamics in this
theory. Houndjo \cite{fRT4} reconstructed $f(R,T)$ gravity by taking
$f(R,T)=f_1(R)+f_2(T)$ and it was proved that $f(R,T)$ gravity
allowed transition of matter from dominated phase to an acceleration
phase. Thus it is hoped that $f(R,T)$ gravity may explain the resent
phase of cosmic acceleration of our universe. This theory can be
used to explore many issues and may provide some satisfactory
results.

In this paper, we are focussed to investigate the exact solutions of
Bianchi types $I$ and $V$ spacetimes in the framework of $f(R,T)$
gravity. The plan of paper is as follows: In section \textbf{2}, we
give some basics of $f(R,T)$ gravity. Section \textbf{3} and
\textbf{4} give the exact solutions for Bianchi types $I$ and $V$
spacetimes. Concluding remarks are given in the last section.

\section{Some Basics of $f(R,T)$ Gravity}

The $f(R,T)$ theory of gravity is the generalization or
modification of GR. The action for this theory is given by
\cite{fRT1}
\begin{equation}\label{frt1}
S=\int\sqrt{-g}\bigg(\frac{1}{16\pi{G}}f(R,T)+L_{m}\bigg)d^4x,
\end{equation}
where $f(R,T)$ is an arbitrary function of the Ricci scalar $R$
and the trace $T$ of energy momentum tensor $T_{\mu\nu}$ while
$L_{m}$ is the usual matter Lagrangian. It is worth mentioning
that if we replace $f(R,T)$ with $f(R)$, we get the action for
$f(R)$ gravity and replacement of $f(R,T)$ with $R$ leads to the
action of GR. The energy momentum tensor $T_{\mu\nu}$ is defined
as \cite{emt}
\begin{equation}\label{frt2}
T_{\mu\nu}=-\frac{2}{\sqrt{-g}}\frac{\delta(\sqrt{-g}L_m)}{\delta
g^{\mu\nu}}.
\end{equation}
Here we assume that the dependance of matter Lagrangian is merely
on the metric tensor $g_{\mu\nu}$ rather than its derivatives. In
this case, we obtain
\begin{equation}\label{frt3}
T_{\mu\nu}=L_m g_{\mu\nu}-2\frac{\delta L_m}{\delta g^{\mu\nu}}.
\end{equation}
The $f(R,T)$ gravity field equations are obtained by varying the
action $S$ Eq.(\ref{frt1}) with respect to the metric tensor
$g_{\mu\nu}$
\begin{equation}\label{frt4}
f_R(R,T)R_{\mu\nu}-\frac{1}{2}f(R,T)g_{\mu\nu}-(\nabla_{\mu}
\nabla_{\nu}-g_{\mu\nu}\Box)f_R(R,T)=\kappa
T_{\mu\nu}-f_T(R,T)(T_{\mu\nu}+\Theta_{\mu\nu}),
\end{equation}
where $\nabla_{\mu}$ denotes the covariant derivative and
\begin{equation*}
\Box\equiv\nabla^{\mu}\nabla_{\mu},~~ f_R(R,T)=\frac{\partial
f_R(R,T)}{\partial R},~~ f_T(R,T)=\frac{\partial
f_R(R,T)}{\partial
T},~~\Theta_{\mu\nu}=g^{\alpha\beta}\frac{\delta
T_{\alpha\beta}}{\delta g^{\mu\nu}}.
\end{equation*}
Contraction of (\ref{frt4}) yields
\begin{equation}\label{frt04}
f_R(R,T)R+3\Box f_R(R,T)-2f(R,T)=\kappa T-f_T(R,T)(T+\Theta),
\end{equation}
where $\Theta={\Theta_\mu}^\mu$. This is an important equation
because it provides a relationship between Ricci scalar $R$ and
the trace $T$ of energy momentum tensor. Using matter Lagrangian
$L_m$, the standard matter energy-momentum tensor is derived as
\begin{equation}\label{frt5}
T_{\mu\nu}=(\rho + p)u_\mu u_\nu-pg_{\mu\nu},
\end{equation}
where $u_\mu=\sqrt{g_{00}}(1,0,0,0)$ is the four-velocity in
co-moving coordinates and $\rho$ and $p$ denotes energy density
and pressure of the fluid respectively. Perfect fluids problems
involving energy density and pressure are not any easy task to
deal with. Moreover, there does not exist any unique definition
for matter Lagrangian. Thus we can assume the matter Lagrangian as
$L_m=-p$ which gives
\begin{equation}\label{frt6}
\Theta_{\mu\nu}=-pg_{\mu\nu}-2T_{\mu\nu},
\end{equation}
and consequently the field equations (\ref{frt4}) take the form
\begin{equation}\label{frt7}
f_R(R,T)R_{\mu\nu}-\frac{1}{2}f(R,T)g_{\mu\nu}-(\nabla_{\mu}
\nabla_{\nu}-g_{\mu\nu}\Box)f_R(R,T)=\kappa
T_{\mu\nu}+f_T(R,T)(T_{\mu\nu}+pg_{\mu\nu}),
\end{equation}
It is mentioned here that these field equations depend on the
physical nature of matter field. Many theoretical models
corresponding to different matter contributions for $f(R,T)$
gravity are possible. However, Harko et al. gave three classes of
these models
\[ f(R,T)= \left\lbrace
  \begin{array}{c l}
    {R+2f(T),}\\
    {f_1(R)+f_2(T),}\\{f_1(R)+f_2(R)f_3(T).}
  \end{array}
\right. \]\\
In this paper we are focussed to the first class, i.e.
$f(R,T)=R+2f(T)$. For this model the field equations become
\begin{equation}\label{frt8}
R_{\mu\nu}-\frac{1}{2}Rg_{\mu\nu}=\kappa
T_{\mu\nu}+2f'(T)T_{\mu\nu}+\bigg[f(T)+2pf'(T)\bigg]g_{\mu\nu},
\end{equation}
where prime represents derivative with respect to $T$.

\section{Exact Solutions of Bianchi Type $I$ Universe}

In this section, we shall find exact solutions of Bianchi I
spacetime in $f(R,T)$ gravity. For this purpose, we use natural
system of units $(G=c=1)$ and $f(T)=\lambda T$, where $\lambda$ is
an arbitrary constant. For Bianchi type $I$ spacetime, the line
element is given by
\begin{equation}\label{6}
ds^{2}=dt^2-A^2(t)dx^2-B^2(t)dy^2-C^2(t)dz^2,
\end{equation}
where $A,~B$ and $C$ are defined as cosmic scale factors. The
Bianchi $I$ Ricci scalar turns out to be
\begin{equation}\label{7}
R=-2\bigg[\frac{\ddot{A}}{A}+\frac{\ddot{B}}{B}+\frac{\ddot{C}}{C}
+\frac{\dot{A}\dot{B}}{AB}+\frac{\dot{B}\dot{C}}{BC}+\frac{\dot{C}\dot{A}}{CA}\bigg],
\end{equation}
where dot denotes derivative with respect to $t$.

Using Eq.(\ref{frt8}), we get four independent field equations,
\begin{eqnarray} \label{11}
\frac{\dot{A}\dot{B}}{AB}+\frac{\dot{B}\dot{C}}{BC}+
\frac{\dot{C}\dot{A}}{CA}=(8\pi+3\lambda)\rho-\lambda
p,\\\label{12} \frac{\ddot{B}}{B}+\frac{\ddot{C}}{C}
+\frac{\dot{B}\dot{C}}{BC}=\lambda
\rho-(8\pi+3\lambda)p,\\\label{13}
\frac{\ddot{C}}{C}+\frac{\ddot{A}}{A}
+\frac{\dot{C}\dot{A}}{AC}=\lambda
\rho-(8\pi+3\lambda)p,\\\label{14}
\frac{\ddot{A}}{A}+\frac{\ddot{B}}{B}
+\frac{\dot{A}\dot{B}}{AB}=\lambda \rho-(8\pi+3\lambda)p.
\end{eqnarray}
These are four non-linear differential equations with five unknowns
namely $A,~B,~C$, $\rho$ and $p$. Subtracting Eq.(\ref{13}) from
Eq.(\ref{12}), Eq.(\ref{14}) from Eq.(\ref{13}) and Eq.(\ref{14})
from Eq.(\ref{11}), we get respectively
\begin{eqnarray}\label{015}
\frac{\ddot{A}}{A}-\frac{\ddot{B}}{B}
+\frac{\dot{C}}{C}\bigg(\frac{\dot{A}}{A}-\frac{\dot{B}}{B}\bigg)=0,\\\label{016}
\frac{\ddot{B}}{B}-\frac{\ddot{C}}{C}
+\frac{\dot{A}}{A}\bigg(\frac{\dot{B}}{B}-\frac{\dot{C}}{C}\bigg)=0,\\\label{017}
\frac{\ddot{A}}{A}-\frac{\ddot{C}}{C}
+\frac{\dot{B}}{B}\bigg(\frac{\dot{A}}{A}-\frac{\dot{C}}{C}\bigg)=0.
\end{eqnarray}
These equations imply that
\begin{eqnarray}\label{15}
\frac{B}{A}=d_1\exp\bigg[{c_1\int\frac{dt}{a^3}}\bigg],\\\label{16}
\frac{C}{B}=d_2\exp\bigg[{c_2\int\frac{dt}{a^3}}\bigg],\\\label{17}
\frac{A}{C}=d_3\exp\bigg[{c_3\int\frac{dt}{a^3}}\bigg],
\end{eqnarray}
where $c_1,~c_2,~c_3$ and $d_1,~d_2,~d_3$ are integration
constants which satisfy the following relation
\begin{equation}\label{18}
c_1+c_2+c_3=0,\quad d_1d_2d_3=1.
\end{equation}
Using Eqs.(\ref{15})-(\ref{17}), we can write the unknown metric
functions in an explicit way
\begin{eqnarray}\label{19}
A=ap_1\exp\bigg[{q_1\int\frac{dt}{a^3}}\bigg],\\\label{20}
B=ap_2\exp\bigg[{q_2\int\frac{dt}{a^3}}\bigg],\\\label{21}
C=ap_3\exp\bigg[{q_3\int\frac{dt}{a^3}}\bigg],
\end{eqnarray}
where
\begin{equation}\label{22}
p_1=({d_1}^{-2}{d_2}^{-1})^{\frac{1}{3}},\quad
p_2=(d_1{d_2}^{-1})^{\frac{1}{3}},\quad
p_3=(d_1{d_2}^2)^{\frac{1}{3}}
\end{equation}
and
\begin{equation}\label{23}
q_1=-\frac{2c_1+c_2}{3},\quad q_2=\frac{c_1-c_2}{3},\quad
q_3=\frac{c_1+2c_2}{3}.
\end{equation}
It is mentioned here that $p_1,~p_2,~p_3$ and $q_1,~q_2,~q_3$ also
satisfy the relation
\begin{equation}\label{24}
p_1p_2p_3=1,\quad q_1+q_2+q_3=0.
\end{equation}

\subsection{Some Important Physical Parameters}

Now we present some important definitions of physical parameters.
The average scale factor $a$ and volume scale factor $V$ are defined
as
\begin{equation}\label{8}
a=\sqrt[3]{ABC}, \quad V=a^3=ABC.
\end{equation}
The generalized mean Hubble parameter $H$ is given by
\begin{equation}\label{008}
H=\frac{1}{3}(H_1+H_2+H_3),
\end{equation}
where
$H_1=\frac{\dot{A}}{A},~H_2=\frac{\dot{B}}{B},~H_3=\frac{\dot{C}}{C}$
are defined as the directional Hubble parameters in the directions
of $x,~y$ and $z$ axis respectively. The mean anisotropy parameter
$A$ is
\begin{equation}\label{0000009}
A=\frac{1}{3}\sum^3_{i=1}\bigg(\frac{H_i-H}{H}\bigg)^2.
\end{equation}
The expansion scalar $\theta$ and shear scalar $\sigma^2$ are
defined as follows
\begin{eqnarray}\label{09}
\theta&=&u^\mu_{;\mu}=\frac{\dot{A}}{A}+\frac{\dot{B}}{B}+\frac{\dot{C}}{C},\\
\label{00009} \sigma^2&=&\frac{1}{2}\sigma_{\mu\nu}\sigma^{\mu\nu}
=\frac{1}{3}\bigg[\bigg(\frac{\dot{A}}{A}\bigg)^2+\bigg(\frac{\dot{B}}{B}\bigg)^2
+\bigg(\frac{\dot{C}}{C}\bigg)^2-\frac{\dot{A}\dot{B}}{AB}-\frac{\dot{B}\dot{C}}{BC}
-\frac{\dot{C}\dot{A}}{CA}\bigg],~~
\end{eqnarray}
where
\begin{equation}\label{009}
\sigma_{\mu\nu}=\frac{1}{2}(u_{\mu;\alpha}h^\alpha_\nu+u_{\nu;\alpha}h^\alpha_\mu)
-\frac{1}{3}\theta h_{\mu\nu},
\end{equation}
$h_{\mu\nu}=g_{\mu\nu}-u_{\mu}u_{\nu}$ is the projection tensor.

The deceleration parameter $q$ is the measure of the cosmic
accelerated expansion of the universe. It is defined as
\begin{equation}\label{26}
q=-\frac{\ddot{a}a}{\dot{a}^2}.
\end{equation}
The behavior of the universe models is determined by the sign of
$q$. The positive value of deceleration parameter suggests a
decelerating model while the negative value indicates inflation.
Since there are four equations and five unknowns, so we use a
well-known relation \cite{15} between the average scale factor $a$
and average Hubble parameter $H$ to solve the equations,
\begin{equation}\label{27}
H=la^{-n},
\end{equation}
where $l$ and $n$ are positive constants.

Using Eqs.(\ref{008}) and (\ref{27}), we get
\begin{equation}\label{28}
\dot{a}=la^{1-n}
\end{equation}
and the deceleration parameter becomes
\begin{equation}\label{29}
q=n-1.
\end{equation}
Integrating Eq.(\ref{28}), it follows that
\begin{equation}\label{30}
a=(nlt+k_1)^{\frac{1}{n}},\quad n\neq0
\end{equation}
and
\begin{equation}\label{31}
a=k_2\exp(lt),\quad n=0,
\end{equation}
where $k_1$ and $k_2$ are constants of integration. Thus we get
two different models of the universe corresponding to these values
of the average scale factor.

\subsection{Singular Model of the Universe}

Here we investigate the model of universe when $n\neq0$, i.e.,
$a=(nlt+k_1)^{\frac{1}{n}}$. In this case, the metric coefficients
$A,~B$ and $C$ takes the form
\begin{eqnarray}\label{35}
A&=&p_1(nlt+k_1)^{\frac{1}{n}}\exp\bigg[\frac{q_1(nlt+k_1)^
{\frac{n-3}{n}}}{l(n-3)}\bigg],\quad n\neq3\\\label{36}
B&=&p_2(nlt+k_1)^{\frac{1}{n}}\exp\bigg[\frac{q_2(nlt+k_1)^
{\frac{n-3}{n}}}{l(n-3)}\bigg],\quad n\neq3\\\label{37}
C&=&p_3(nlt+k_1)^{\frac{1}{n}}\exp\bigg[\frac{q_3(nlt+k_1)^
{\frac{n-3}{n}}}{l(n-3)}\bigg],\quad n\neq3.
\end{eqnarray}
The directional Hubble parameters $H_i$ ($i=1,2,3$) turn out to be
\begin{equation}\label{38}
H_i=\frac{l}{nlt+k_1}+\frac{q_i}{(nlt+k_1)^{\frac{3}{n}}}.
\end{equation}
The mean generalized Hubble parameter and volume scale factor are
\begin{equation}\label{39}
H=\frac{l}{nlt+k_1},\quad V=(nlt+k_1)^\frac{3}{n}.
\end{equation}
The mean anisotropy parameter becomes
\begin{equation}\label{3929}
A=\frac{{q_1}^2+{q_2}^2+{q_3}^2}{3l^2(nlt+k_1)^{(6-2n)/n}}.
\end{equation}
The expansion scalar and shear scalar for this model are given by
\begin{equation}
\theta=\frac{3l}{nlt+k_1},\quad\sigma^2=\frac{{q_1}^2+{q_2}^2+{q_3}^2}{2(nlt+k_1)^{6/n}}.
\end{equation}
Using Eqs. (\ref{11})-(\ref{14}), the energy density of the
universe is
\begin{eqnarray}\nonumber
\rho=&&\frac{1}{12(\lambda+2\pi)(\lambda+4\pi)}\bigg[4(\lambda+3\pi)
\bigg\{\frac{3l^2}{(nlt+k_1)^2}+\frac{q_1q_2+q_2q_3+q_3q_1}{(nlt+k_1)^{\frac{6}{n}}}\bigg\}\\\label{ro1}
&-&\lambda
\bigg\{\frac{3l^2(1-n)}{(nlt+k_1)^2}+\frac{{q_1}^2+{q_2}^2+{q_3}^2}{(nlt+k_1)^{\frac{6}{n}}}\bigg\}\bigg]
\end{eqnarray}
while the pressure of the universe becomes
\begin{eqnarray}\nonumber
p=&&\frac{-1}{12(\lambda+2\pi)(\lambda+4\pi)}\bigg[4\pi
\bigg\{\frac{3l^2}{(nlt+k_1)^2}+\frac{q_1q_2+q_2q_3+q_3q_1}{(nlt+k_1)^{\frac{6}{n}}}\bigg\}\\\label{p1}
&+&(3\lambda+8\pi)\bigg\{\frac{3l^2(1-n)}{(nlt+k_1)^2}+\frac{{q_1}^2+{q_2}^2+{q_3}^2}{(nlt+k_1)^{\frac{6}{n}}}\bigg\}\bigg].
\end{eqnarray}

\subsection{Non-singular Model of the Universe}

For this model, $n=0$ and the average scale factor $a=k_2\exp(lt)$
turns the metric coefficients $A,~B$ and $C$ into
\begin{eqnarray}\label{43}
A&=&p_1k_2\exp(lt)\exp\bigg[-\frac{q_1\exp(-3lt)}{3l{k_2}^3}\bigg],\\\label{44}
B&=&p_2k_2\exp(lt)\exp\bigg[-\frac{q_2\exp(-3lt)}{3l{k_2}^3}\bigg],\\\label{37}
C&=&p_3k_2\exp(lt)\exp\bigg[-\frac{q_3\exp(-3lt)}{3l{k_2}^3}\bigg].
\end{eqnarray}
The directional Hubble parameters $H_i$ become
\begin{equation}\label{44}
H_i=l+\frac{q_i}{{k_2}^3}\exp(-3lt) .
\end{equation}
The mean generalized Hubble parameter and volume scale factor
turns out to be
\begin{equation}\label{46}
H=l,\quad V={k_2}^3\exp(3lt).
\end{equation}
The mean anisotropy parameter, expansion scalar and shear scalar
are
\begin{equation}
A=\frac{{q_1}^2+{q_2}^2+{q_3}^2}{3l^2{k_2}^6\exp(6lt)},~~
\theta=3l,\quad
\sigma^2=\frac{{q_1}^2+{q_2}^2+{q_3}^2}{2{k_2}^6\exp(6lt)}.
\end{equation}
For this model, the energy density and pressure of the universe
takes the form
\begin{eqnarray}\nonumber
\rho=&&\frac{1}{12(\lambda+2\pi)(\lambda+4\pi)}\bigg[4(\lambda+3\pi)
\bigg\{3l^2+\frac{q_1q_2+q_2q_3+q_3q_1}{{k_2}^6\exp(6lt)}\bigg\}\\\label{ro2}
&-&\lambda\bigg\{3l^2+\frac{{q_1}^2+{q_2}^2+{q_3}^2}{{k_2}^6\exp(6lt)}\bigg\}\bigg],
\end{eqnarray}
\begin{eqnarray}\nonumber
p=&&\frac{-1}{12(\lambda+2\pi)(\lambda+4\pi)}\bigg[4\pi
\bigg\{3l^2+\frac{q_1q_2+q_2q_3+q_3q_1}{{k_2}^6\exp(6lt)}\bigg\}\\\label{p2}
&+&(3\lambda+8\pi)\bigg\{3l^2+\frac{q_1q_2+q_2q_3+q_3q_1}{{k_2}^6\exp(6lt)}\bigg\}\bigg].
\end{eqnarray}

\section{Exact Bianchi Type $V$ Solutions}

Here we shall explore Bianchi type $V$ solutions in the context of
$f(R,T)$ gravity. The metric for the Bianchi type $V$ spacetime is
\begin{equation}\label{56}
ds^{2}=dt^2-A^2(t)dx^2-e^{2mx}[B^2(t)dy^2+C^2(t)dz^2].
\end{equation}
Here $A,~B$ and $C$ are also cosmic scale factors and $m$ is an
any constant. The Ricci scalar for this spacetime is
\begin{equation}\label{57}
R=-2\bigg[\frac{\ddot{A}}{A}+\frac{\ddot{B}}{B}+\frac{\ddot{C}}{C}-\frac{3m^2}{A^2}
+\frac{\dot{A}\dot{B}}{AB}+\frac{\dot{B}\dot{C}}{BC}+\frac{\dot{C}\dot{A}}{CA}\bigg].
\end{equation}
Using Eq.(\ref{frt8}), we get
\begin{eqnarray} \label{511}
\frac{\dot{A}\dot{B}}{AB}+\frac{\dot{B}\dot{C}}{BC}+
\frac{\dot{C}\dot{A}}{CA}-\frac{3m^2}{A^2}=(8\pi+3\lambda)\rho-\lambda
p,\\\label{512} \frac{\ddot{B}}{B}+\frac{\ddot{C}}{C}
+\frac{\dot{B}\dot{C}}{BC}-\frac{m^2}{A^2}=\lambda
\rho-(8\pi+3\lambda)p,\\\label{513}
\frac{\ddot{C}}{C}+\frac{\ddot{A}}{A}
+\frac{\dot{C}\dot{A}}{AC}-\frac{m^2}{A^2}=\lambda
\rho-(8\pi+3\lambda)p,\\\label{514}
\frac{\ddot{A}}{A}+\frac{\ddot{B}}{B}
+\frac{\dot{A}\dot{B}}{AB}-\frac{m^2}{A^2}=\lambda
\rho-(8\pi+3\lambda)p
\end{eqnarray}
and the $01$-component turn out to be
\begin{equation}\label{0513}
2\frac{\dot{A}}{A}-\frac{\dot{B}}{B}-\frac{\dot{C}}{C}=0.
\end{equation}
We adopt the same procedure as for the Bianchi type $I$ solutions to
solve these equations. Here the equations
Eqs.(\ref{015})-(\ref{017}) are same as obtained previously but by
making use of Eq.(\ref{0513}), we get the constraint equations as
follows
\begin{equation}\label{0524}
p_1=1,\quad p_2={p_3}^{-1}=P,\quad q_1=0,\quad q_2=-q_3=Q.
\end{equation}
Thus, the metric coefficients become
\begin{eqnarray}\label{0519}
A=a,\quad B=aP\exp\bigg[{Q\int\frac{dt}{a^3}}\bigg],\quad
C=aP^{-1}\exp\bigg[{-Q\int\frac{dt}{a^3}}\bigg].
\end{eqnarray}

\subsection{Singular Model of the Universe}

For the model of the universe when $n\neq0$, the metric functions
$A,~B$ and $C$ become
\begin{eqnarray}\label{535}
A&=&(nlt+k_1)^{\frac{1}{n}},\\\label{536}
B&=&P(nlt+k_1)^{\frac{1}{n}}\exp\bigg[\frac{Q(nlt+k_1)^
{\frac{n-3}{n}}}{l(n-3)}\bigg],\quad n\neq3\\\label{537}
C&=&P^{-1}(nlt+k_1)^{\frac{1}{n}}\exp\bigg[\frac{-Q(nlt+k_1)^
{\frac{n-3}{n}}}{l(n-3)}\bigg],\quad n\neq3.
\end{eqnarray}
The directional Hubble parameters $H_1,~H_2$ and $H_3$ take the
form
\begin{eqnarray}\label{538}
H_1&=&\frac{l}{nlt+k_1},\\\label{0538}
H_2&=&\frac{l}{nlt+k_1}+\frac{Q}{(nlt+k_1)^{\frac{3}{n}}},\\\label{00538}
H_2&=&\frac{l}{nlt+k_1}-\frac{Q}{(nlt+k_1)^{\frac{3}{n}}}.
\end{eqnarray}
The mean anisotropy parameter becomes
\begin{equation}\label{3929}
A=\frac{2Q^2}{3(nlt+k_1)^{(6-2n)/n}}.
\end{equation}
The shear scalar for this model is given by
\begin{equation}
\sigma^2=\frac{Q^2}{(nlt+k_1)^{6/n}}.
\end{equation}
It is to be noticed that the mean generalized Hubble parameter
$H$, expansion scalar $\theta$ and the volume scale factor $V$
are same as in the case of Bianchi type $I$ spacetime. The energy
density and pressure of Bianchi $V$ universe for this model turns
out to be
\begin{eqnarray}\nonumber
\rho=&&\frac{1}{12(\lambda+2\pi)(\lambda+4\pi)}\bigg[4(\lambda+3\pi)
\bigg\{\frac{3l^2}{(nlt+k_1)^2}-\frac{Q^2}{(nlt+k_1)^{\frac{6}{n}}}\bigg\}\\\label{ro3}
&-&\lambda\bigg\{\frac{3l^2(1-n)}{(nlt+k_1)^2}+\frac{2Q^2}{(nlt+k_1)^{\frac{6}{n}}}\bigg\}\bigg],
\end{eqnarray}
\begin{eqnarray}\nonumber
p=&&\frac{-1}{12(\lambda+2\pi)(\lambda+4\pi)}\bigg[4\pi
\bigg\{\frac{3l^2}{(nlt+k_1)^2}-\frac{Q^2}{(nlt+k_1)^{\frac{6}{n}}}\bigg\}\\\label{p3}
&+&(3\lambda+8\pi)\bigg\{\frac{3l^2(1-n)}{(nlt+k_1)^2}+\frac{2Q^2}{(nlt+k_1)^{\frac{6}{n}}}\bigg\}\bigg].
\end{eqnarray}

\subsection{Non-singular Model of the Universe}

For the model when $n=0$, the metric coefficients $A,~B$ and $C$
turn out to be
\begin{eqnarray}\label{543}
A&=&k_2\exp(lt),\\\label{544}
B&=&Pk_2\exp(lt)\exp\bigg[-\frac{Q\exp(-3lt)}{3l{k_2}^3}\bigg],\\\label{545}
C&=&P^{-1}k_2\exp(lt)\exp\bigg[\frac{Q\exp(-3lt)}{3l{k_2}^3}\bigg].
\end{eqnarray}
The directional Hubble parameters $H_1,~H_2$ and $H_3$ are
\begin{eqnarray}\label{546}
H_1=l,\quad H_2=l+\frac{Q\exp(-3lt)}{{k_2}^3},\quad
H_3=l-\frac{Q\exp(-3lt)}{{k_2}^3}.
\end{eqnarray}
The mean anisotropy parameter and shear scalar for this model
become
\begin{equation}
A=\frac{2Q^2}{3l^2{k_2}^6\exp(6lt)},~~~~\quad
\sigma^2=\frac{Q^2}{{k_2}^6\exp(6lt)}.
\end{equation}
Here we also get the same volume scale factor $V$, expansion
scalar $\theta$ and mean generalized Hubble parameter $H$ as shown
in Eqs.(\ref{44})-(\ref{46}). The energy density and pressure of
universe here become
\begin{eqnarray}\nonumber
\rho=&&\frac{1}{12(\lambda+2\pi)(\lambda+4\pi)}\bigg[4(\lambda+3\pi)
\bigg\{3l^2-\frac{Q^2}{{k_2}^6\exp(6lt)}\bigg\}\\\label{ro4}
&-&\lambda\bigg\{3l^2+\frac{2Q^2}{{k_2}^6\exp(6lt)}\bigg\}\bigg],
\end{eqnarray}
\begin{eqnarray}\nonumber
p=&&\frac{-1}{12(\lambda+2\pi)(\lambda+4\pi)}\bigg[4\pi
\bigg\{3l^2-\frac{Q^2}{{k_2}^6\exp(6lt)}\bigg\}\\\label{p4}
&+&(3\lambda+8\pi)\bigg\{3l^2+\frac{2Q^2}{{k_2}^6\exp(6lt)}\bigg\}\bigg].
\end{eqnarray}

\section{Concluding Remarks}

This paper is devoted to discuss the current phenomenon of
accelerated expansion of universe in the framework of newly
proposed $f(R,T)$ theory of gravity. For this purpose, we take
$f(R,T)=R+2\lambda T$ and explore the exact solutions of Bianchi
types $I$ and $V$ cosmological models. We obtain two exact
solutions for both spacetimes using the assumption of constant
value of deceleration parameter and the law of variation of Hubble
parameter. The obtained solutions correspond to two different
models of universe. The first solution forms a singular model with
power law expansion while the second solution gives a non-singular
model with exponential expansion of universe. The physical
parameters for both of these models are discussed below.

The singular model of the universe corresponds to $n\neq0$ with
average scale factor $a=(nlt+k_1)^{\frac{1}{n}}$. This model
possesses a point singularity when $t\equiv t_s=-\frac{k_1}{nl}$.
The volume scale factor vanishes and the metric coefficients
$A,~B$ and $C$ vanish at this singularity point. The cosmological
parameters $H_1,~H_2,~H_3,~H,~ \theta$, and $\sigma^2$ are all
infinite at this point of singularity. The mean anisotropy
parameter $A$ also becomes infinite at this point for $0<n<3$ and
vanishes for $n>3$. The energy density and pressure of universe
are also infinite at this epoch. Moreover, the isotropy condition,
i.e., $\frac{\sigma^2}{\theta}\rightarrow 0$ as $t\rightarrow
\infty$, is verified for this model. All these conclusive
observations suggest that the universe starts its expansion with
zero volume, strong pressure and energy density from $t=t_s$ and
it will continue to expand for $0<n<3$.

Now we discuss the non-singular model of the universe corresponds
to $n=0$. For this model the average scale factor is
$a=k_2\exp(lt)$. The non-singularity is due to the exponential
behavior of the model. The expansion scalar $~\theta$ and mean
generalized Hubble parameter $H$ are constant in this case. For
finite values of $t$, the physical parameters
$H_1,~H_2,~H_3,~\sigma^2$ and $A$ are all finite. The metric
functions are defined for finite time and the isotropy condition
is satisfied. The pressure and energy density of universe become
infinite in the limiting case when $t\rightarrow -\infty$. This
shows that the universe evolved from an infinite past with a
massive energy density and pressure. There is an exponential
increase in the volume as the time grows. This shows that the
universe started its expansion a long time ago with zero volume.

\end{document}